# Effects of Head-locked Augmented Reality on User's Performance and Perceived Workload

Yalda Ghasemi, Ankit Singh, Myunghee Kim, Andrew Johnson, and Heejin Jeong
University of Illinois at Chicago

An augmented reality (AR) environment includes a set of digital elements with which the users interact while performing certain tasks. Recent AR head-mounted displays allow users to select how these elements are presented. However, few studies have been conducted to examine the effect of presenting augmented content on user performance and workload. This study aims to evaluate two methods of presenting augmented content — world-locked and head-locked modes in a data entry task. A total of eighteen participants performed the data entry task in this study. The effectiveness of each mode is evaluated in terms of task performance, muscle activity, perceived workload, and usability. The results show that the task completion time is shorter, and the typing speed is significantly faster in the head-locked mode while the world-locked mode achieved higher scores in terms of preference. The findings of this study can be applied to AR user interfaces to improve content presentation and enhance the user experience.

## INTRODUCTION

Augmented reality (AR) has increasingly gained popularity over the past few decades. AR is an interactive experience of the real world superimposing computer-generated elements as virtual information. In other words, AR transforms reality by adding digital elements to the real environment. These digital elements may include images, videos, global positioning system annotations, or other elements that do not exist in the real environment. However, AR provides the ability to interact with these elements in the real environment in real-time.

AR technology allows several tasks to be performed through its user interfaces and natural interactions (Carmigniani et al., 2011). AR has been used in many areas to assist workers by making tasks more convenient and interactive. Displaying information is an essential part of using AR. Information in the AR environment should be displayed to the users properly so that they would be able to interact with the content in the most efficient way. There are generally two ways of presenting information in AR head-mounted displays (HMDs), including static view and following view. In the static view, also referred to as world-locked, the AR content is placed in a fixed position in the environment, and its position does not change with the user's movements. In this mode, the AR system recognizes the user's environment to overlay the content. The overlaid content stays in the environment even if the user walks around. When there is enough light in an environment with a clear texture and the sensors of the HMD are not blocked, this method can be applied successfully. The static view is usually used as the default mode in HMDs. On the other hand, the following view is more flexible and can be adjusted automatically based on the user's movements. This method can be divided into two modes, i.e., head-locked and body-locked. In the head-locked mode, the augmented content moves based on the user's head movements, and it is always in the user's field of view (FOV). In this mode, the position and rotation of the content presented to the user change exactly according to the camera motion. The content always follows the user's head movement and stays at the same distance from the camera. In the body-locked mode, the content moves with respect to the user's body movements and position.

In this study, a data entry task was used to explore the utility of head-locked content for data presentation in an AR environment (Singh et al., 2020). The world-locked content or static view was also included and considered as the baseline. The data entry task is defined as a task where the operator receives the required information on a screen or paper and enters the presented information into a target device (e.g., a computer).

This study investigates which one of these two methods is a better choice for displaying information in AR environments by evaluating their effects on the user performance and perceived workload of the task. In addition, the electromyogram (EMG) test was used to measure the muscular activity of the neck and shoulders for both methods.

## RELATED WORK

Presenting information in AR has applications in many areas, such as education (Guntur et al., 2020) when students are experiencing an engaging classroom with overlaid digital elements in the environment or the industrial tasks (Malik et al., 2020) for collaborative tasks between humans and robots. The overlaid AR content could be fixed in the first displayed position or move with the user's movements. One of the first studies on the evaluation of information displaying in wearable technologies was conducted by Bilinghurst et al. (1998). In their study, the effects of head-stabilized and body-stabilized displays were compared. In the head-stabilized method, the content is fixed to the users' viewpoint and does not change as the users change their position and orientation. However, in the body-stabilized method, the content is fixed with respect to the users' body position and changes only based on the users' viewpoint orientation. The study suggested that the users' body-stabilized method resulted in better performance, especially in the search tasks.

The impact of cybersickness or virtual reality-induced symptoms and effects (VRISE) was investigated for rotational motions of the AR/VR interface (Kemeny et al., 2017). A new head-lock navigation method for rotational motions in a virtual environment was developed to generate a "Pseudo AR" mode through this study. Further experiments demonstrated that this novel navigation technique reduced VRISE in participants significantly. Klose et al. (2019) proposed a text presentation for walking dual tasks using both body-locked and head-locked content. The results of this study showed that the users preferred the head-locked content. It was also found that the body-locked content was more distracting for the users. Fiannaca et al. (2014) developed a head-locked navigation app for optical HMDs. This system is beneficial for blind people to assist them by providing feedback when a landmark is detected. In this study, the head-locked content was mainly designed to prevent users from veering. Fukushima et al. (2020) showed that text reading speed in an optical see-through HMD was significantly increased when using head-locked content. On the other hand, the Magic Leap AR device developers have stated that augmenting objects in a 3-dimensional spatial layout and having them following head movement with differing amounts of lag negatively affects the user experience (Magic Leap Developer, 2020). Also, Microsoft discouraged using head-locked content by stating that this method could be unnatural and uncomfortable for the users (Turner & Coulter, 2019). Although the head-locked content was the subject of many studies, there were not enough studies showing the difference between head-locked and world-locked modes and comparing whether they can affect the user performance from cognitive to physical aspects.

This study aims to reduce the amount of head movement involved in accessing the data from the data presentation interface and entering that data into the system. Since the data entry task often causes physical and cognitive fatigue in prolonged durations of working (Healy et al., 2014), it is vital to address the efficiency of information presentation methods when performing the AR task. To this end, we test the performance of the head-locked content for a seated data entry task and investigate how it affects participants' performance compared to the world-locked mode. Having the image follow the head movement could significantly reduce the head movement's up and down (or sideward) movements involved in accessing the information by presenting the data to the user's FOV (Jeong et al., 2020).

The users' electromyography was measured to identify which of the two view modes has less muscular activity and less effect on the user fatigue. In addition, the usability and perceived workload of both methods were evaluated along with their associated accuracy, typing speed, and task duration. We hypothesized that in a data entry task:

[H1] Head-locked method reduces workload and has better usability metrics for users compared to the head-lock view mode.
[H2] Head-locked method improves accuracy, typing speed, and task duration for the data entry task compared to the head-locked view mode.
[H3] Head-locked method involves less muscular fatigue in the neck and shoulders than the world-locked view mode.

## METHOD

To test the hypotheses, a data entry task was used to evaluate the AR content presentation in two view modes of world-locked and head-locked. An illustration of the two modes is presented in Figure 1.

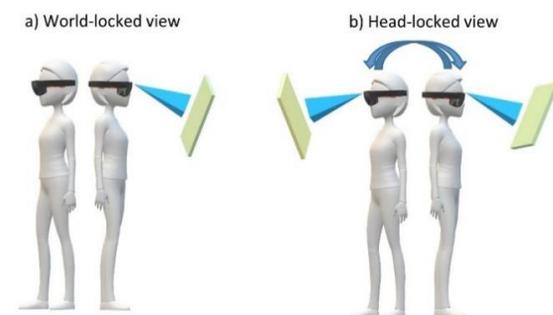

Figure 1. World-locked and head-locked view modes

In this study, a total of 18 participants consisting of 14 males and 4 females were recruited for the experiment (mean age = 21.9, $SD$ = 1.87). The demographic data showed that 5 out of 18 participants had no prior experience using AR HMD. All participants had normal vision, and none of them wore glasses since this might negatively affect their performance on using the AR device. Images of two forms (e.g., US Passport Application and University Faculty Advising) were used for displaying in the AR environment. A counterbalanced experiment was designed to prevent any bias toward the order of the presentation methods and type of the forms. All of the participants experienced each condition (i.e., 4 trials). There was no time limit specified, and the participants were asked to complete each trial at their own pace.

The Magic Leap One HMD was used to present the AR content to the users. The users were asked to enter the observed information from the AR device into a computer. The users were trained before the experiment to become familiar with the interface. While doing the task, a wireless EMG device, Delsys Tringo™, was used to measure the muscular activity of the neck and shoulders. Based on a study by Mclean and Urquhart (2002), the upper trapezius muscle is the most affected area of the neck and upper shoulder in a data entry task. Therefore, the EMG electrodes were attached to this area with a 5 – 7 cm distance to measure the muscular activity in the right and left upper trapezius.

The EMG data collection was done at a rate of 1260 HZ, and the signals were stored on the computer using EMGworks® software. After finishing each trial, participants were asked to complete the NASA-TLX (Hart & Staveland, 1988) for rating their perceived workload on a scale of 0 – 100, along with a customized survey for rating the usability of the corresponding modes on a scale of 1 – 7. The survey questions are presented in Table 1. Finally, the participants were asked to share their feedback and suggestions for each mode after filling out the questionnaires.

Table 1. Usability Metrics Questionnaire

| Mode | Questions |
|---|---|
| World-locked (static) | The use of the AR static interface was helpful for data entry tasks. |
| | The use of the AR static interface was easy for me to perform data entry tasks. |
| | I prefer the use of the AR static interface for data entry tasks. |
| Head-locked (following) | I prefer the use of the AR following interface for data entry tasks. |
| | The use of the AR following interface was easy for me to perform data entry tasks. |
| | I prefer the use of the AR following interface for data entry tasks. |

In order to conduct the data entry task, the participants were instructed to wear the AR device that augmented a scanned image in the environment and enter the data into the data entry device. The size of the augmented content was fitted to the user's FOV in the HMD, and they were able to see the entire length of the image without scrolling the content. Figure 2 shows the user performing the data entry task using the Magic Leap headset.

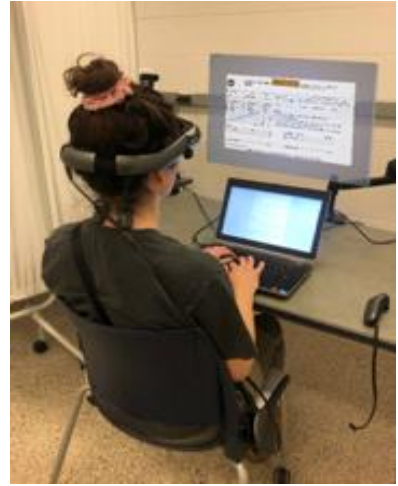

Figure 2. The AR interface

*World-locked (static view)*: In the world-locked content mode, the images of the forms were positioned in the center, above the data entry device. The position of the scanned image was fixed on the AR display concerning the surroundings. The participant had to look directly at the image to obtain the information. The image was presented only in the display frame for the surrounding background of the room. If the participants moved their heads (vertically or horizontally), the information would disappear or be partially visible. The entirety of the document was visible at one time in the AR interface, and the participants did not need to move their heads to scan through the different parts of the document.

*Head-locked (following view)*: In the head-locked content mode, although the position of the scanned image was fixed on the AR display at the same distance as of the static view mode, the image would move along with the participants' head movements. The position of the image was augmented to the user's surroundings. It was not fixed to the background, and the participants could actually have the image follow them to different frames with their head movements.

## RESULTS

In this section, the study results in terms of workload, usability, muscular activity, typing speed, error rate, and task duration are presented. A selection of the participants' feedback and suggestions are also reported.

**Perceived Workload (NASA-TLX)**

Based on the results of the statistical analysis for the NASA-TLX questionnaire, it was found that the two modes of displaying AR content did not show any significant difference in their scores (p-value > .05 for the Mann-Whitney U test) in terms of subjective workload. In addition to the individual workload, the pairwise importance of the six subscales involved in the NASA-TLX survey and creating a weighted rating of those factors based on the importance of each factor as stated by the participants, it could be seen that they also had no significant difference in their scores (p-value > .05 for the Mann-Whitney U test). These results do not support the hypothesis [H1]. Therefore, the head-locked method has no significant effect on the physical and mental workloads for the data entry task compared to the world-locked method.

**Usability - Helpfulness, Ease of Use, and Preference**

The results of statistical analysis of the usability questionnaire showed that participants rated the world-locked mode higher than the head-locked mode in terms of preference (Mean Rank for the Mann-Whitney U test 41.71 > 29.29, p-value = .009) (Fig. 3(a)). However, the results indicated that Helpfulness and Ease of Use did not show any significant difference in their scores from each other (p-value > .05 for the Mann-Whitney U test). Therefore, these results do not support the hypothesis [H1]. It can be stated that the head-locked view mode does not significantly reduce the physical and mental workloads for the data entry task compared to the world-locked method.

**Typing Speed and Error Rate**

It was found that participants' typing speed was higher with the head-locked content compared to the world-locked content (Mean Rank for the Mann-Whitney U test 42.40 > 27.81, p-value = .003) (Fig. 3(b)). This result supports the hypothesis [H2]. Therefore, the head-locked content improves typing speed for data entry tasks. The error rate was not significantly different for the two modes (p-value > .05 for the Mann-Whitney U test). This result does not support the hypothesis [H2]. Therefore, the head-locked content does not significantly improve accuracy for the data entry task compared to the world-locked content.

**Task Duration**

Statistical analysis showed that the time taken to complete the data entry task was higher when using the world-locked mode than it was when using the head-locked mode (Mean value for the t-test 133.3 > 110.5, p-value = .006) (Fig 3(c)). This result supports the hypothesis [H2]. It can be stated that the head-locked content significantly reduces task duration for the data entry task compared to the world-locked mode.

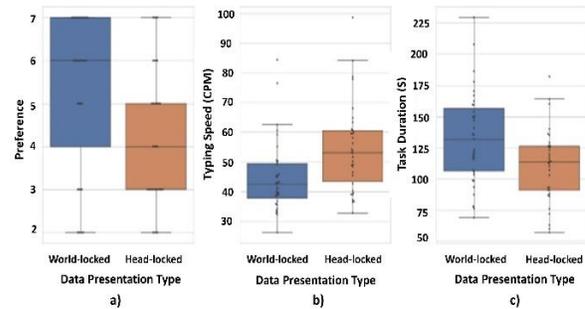

Figure 3. Results of usability, typing speed, and task duration in two view modes

**EMG Analysis**

The results of the statistical analysis of EMG data showed no significant difference between the two modes in the case of muscle activity measured as %MVC (p-value > .05 for the Mann-Whitney U test). This result does not support the hypothesis [H3]. Therefore, the head-locked content has no significant effect on the physical workload for the data entry task compared to the world-locked content.

**Participants' Feedback and Suggestions**

Participants made suggestions through the open-ended survey provided to them at the end of the study to improve the current version of the AR interface for the two modes. The suggestions made by the participants include 1) improving the motion sensitivity to the head movements for the head-locked mode. 2) it was stated that the head-locked mode was more convenient to use than the static view mode. 3) developing an interface containing a zoom option was also suggested for this type of task.

## DISCUSSION

When comparing the head-locked and world-locked content in a data entry task, the participants perceived the similar workload for the two modes. This can be the result of a general lack of experience in using AR devices. Since both modes were equally new to the participants, comprehending and working with the AR device may require considerable effort. Furthermore, participants had a higher typing speed in the head-locked mode. In this mode, the information was available to them at all angles and levels, and they did not spend additional time in the process of looking up to a fixed point above the monitor where the data were displayed. This consequently decreased the task duration for performing the data entry using the head-locked mode, but its effect was not significant on the physical load of the neck and shoulders. Although developers discouraged using head-locked content, this study showed that this method cannot be necessarily inferior to the world-locked or static view mode. The head-locked method resulted in a shorter task completion time and higher typing speed than the world-locked mode. While most of the users preferred the world-locked content in terms of preference, there was no significant difference between the two modes in terms of other measures.

## CONCLUSION

In this study, the effectiveness of two AR presentation methods including head-locked and world-locked content has been evaluated in a data entry task. Based on the results, the head-locked content displayed better performance in task completion time and typing speed while the conventional static view mode achieved higher scores in terms of preference. The participants suggested introducing a zoomable user interface and enhancing the motion sensitivity for the head-locked mode. Since there are various applications of using AR, the selection of presentation methods should be made according to the purpose of the task. For the seated data entry task tested in this study, the results showed that the two modes do not have significant differences in workload and performance.